\documentstyle[preprint,eqsecnum,aps]{revtex}
\begin{document}
%
\begin{titlepage}
\preprint{UCSBTH-95-28}
\title{Strong Decoherence\thanks{Contribution to the {\sl Proceedings of
the 4th Drexel Symposium on Quantum Non-Integrability --- The Quantum-Classical
Correspondence}, Drexel University, September 8-11, 1994}}

\tighten
\author{Murray Gell-Mann}
\address{Santa Fe Institute, 1399 Hyde Park Rd\\
Santa Fe, NM 87501\\}
\address{Los Alamos National Laboratory, MS-B210\\
Los Alamos, NM 87545}
\address{Department of Physics \& Astronomy, University of New Mexico\\
Albuquerque, NM 87131}
\author{and}
\author{James B.~Hartle\thanks{e-mail address:
hartle@cosmic.physics.ucsb.edu}}
\address{Department of Physics, University of California\\
Santa Barbara, CA 93106-9530
}
\date{\today}

\maketitle

\begin{abstract}
\tighten

We introduce a condition for the strong decoherence of a set of
alternative histories of a closed quantum-mechanical system such as the
universe. The condition applies, for a pure initial state,
 to sets of homogeneous histories that
are chains of projections, generally branch-dependent.
Strong decoherence implies the consistency of
probability sum rules but not every set of consistent or even medium
decoherent histories is strongly decoherent.  Two conditions
characterize a strongly decoherent set of histories: (1) At any time the
operators that effectively commute with generalized records of history
up to that moment provide the pool from which --- with suitable adjustment
for elapsed time --- the chains of projections
extending history to the future may be drawn. (2) Under the adjustment
process, generalized record operators acting on the initial state of the
universe are approximately unchanged. This expresses the permanence of
generalized records. The strong decoherence conditions (1) and (2)
guarantee what we call ``permanence of the past'' --- in particular
the continued decoherence of past alternatives as the chains of
projections are extended into the future. Strong
decoherence is an idealization capturing in
a general way this and other aspects of
realistic physical mechanisms that destroy interference, as we illustrate
in a simple model.  We discuss the connection between the reduced density
matrices that have often been used to characterize mechanisms of decoherence
and the more general notion of strong decoherence. The relation between
strong decoherence and a measure of
classicality is briefly described.

\end{abstract}
\end{titlepage}

\tighten

\section{Introduction}
\label{sec:I}

In this article we continue our efforts to explore the quantum mechanics of
closed
systems, most generally and realistically the universe as a whole,
and within that framework to understand the significance of the quasiclassical
realm\footnote{\setlength{\baselineskip}{.1in} In previous work
we called a decoherent set of alternative
coarse-grained histories a ``domain''. However, that term can be confusing
because of its
other uses in physics. We do not want to call such a set a
``world'' because that word connotes a single history and not a set of
alternative ones. Hence, we now call a decoherent set of alternative
coarse-grained histories a ``realm''.} that includes
familiar experience.

We introduce a strong realistic
principle of decoherence for sets of alternative coarse-grained histories of a
closed
 system and
discuss the relationships between this and other principles that have been put
forward. We also examine the concept of classicality, including the role
played by the realistic principle of decoherence in characterizing it.

The most general predictions of quantum mechanics are the
probabilities of individual members of a set of alternative coarse-grained
histories of the universe.\footnote{\setlength{\baselineskip}{.1in}
These are {\it a priori} probabilities.
They can also be thought of as the statistical probabilities for an ensemble
of universes, but in that case we have access only to one member of this
ensemble.}
A set of
coarse-grained histories is a partition of one of the sets of
fine-grained histories
 (which are the most refined  possible
descriptions of the closed system) into mutually exclusive classes. The
classes are the individual coarse-grained histories, which are thus ``bundles''
of fine-grained histories.

The absence of quantum-mechanical interference between the
individual histories in a set is necessary, at the very least,
 for quantum theory to assign
consistent probabilities to the alternative possibilities.
Such sets of histories for which interference is absent
are said to {\it decohere}. Except for pathological
cases, coarse-graining is necessary for decoherence.

Various conditions for the  decoherence of sets of histories have been
proposed.
Some authors have tried to weaken the condition as much as possible to get the
minimum condition necessary for probabilities to be defined. Our point of view
has
always been to try and describe a realistic principle of decoherence that
characterizes in a general way
 the physical processes by which the dissipation of
interference occurs. We have therefore been led to investigate conditions of
decoherence that are as strong as possible compatible with the physical
mechanisms
that destroy interference. We shall investigate such a strong
condition for decoherence in this paper.

Implementing a strong form of decoherence is part of a program to understand
how the quasiclassical realm that includes everyday experience arises in
quantum mechanics
from the Hamiltonian of the elementary particles and the initial condition of
the universe \cite{GH90a}. By a quasiclassical realm we mean an exhaustive
set of mutually
exclusive coarse-grained alternative histories that obey a realistic principle
of decoherence, that consist largely of similar but
branch-dependent alternatives
at a succession of times, with individual histories exhibiting patterns of
correlations implied by effective classical equations of motion subject to
frequent
small fluctuations and occasional major ones, the whole set being maximally
refined given these properties. The theory may exhibit
essentially inequivalent quasiclassical realms,
but there is certainly at least one that includes familiar experience.
This is the {\it usual} quasiclassical realm, described, at least in part, by
alternative values of
hydrodynamic operators that are integrals, over suitable volumes, of densities
of
conserved or nearly conserved quantities.
Examples are densities of energy, momentum, baryon number, and, in late epochs
of the universe, of
nuclei and even chemical species. The sizes of the volumes and
the spacing of the alternatives in time are limited above by maximality.
The size and spacing are limited below by decoherence and the requirement that
the volumes have sufficient ``inertia'' to enable them to resist deviations
from predictability caused by quantum spreading and by the noise that typical
mechanisms
of decoherence produce \cite{GH90a,GH93a}.

A key property of the usual quasiclassical realm is the persistence of
the past. Histories of quasiclassical alternatives up to a given time
can be extended into the future to give  further such histories
without endangering the decoherence of the past
alternatives.
This persistence of the past is not guaranteed by quantum mechanics alone.
Extending  a set of histories into the future is a kind of fine graining and
this
carries the risk of  losing
decoherence.\footnote{\setlength{\baselineskip}{.1in}Indeed, Dowker and Kent
\cite{DKpp} have given
examples with special final conditions where a quasiclassical realm cannot be
extended at all.}  However, the persistence of the past is critical to the
utility
of the quasiclassical realm. It is the reason that we do not need to do an
elaborate
calculation verifying the preservation of
 past decoherence on every occasion when we want to predict the
probability of a quasiclassical alternative in the future conditioned on
our experience of the past. We proceed, secure in the understanding that in
a quasiclassical realm the past (including the decoherence of past
alternatives)
will continue  to persist.

In this article we discuss a strong form of decoherence that guarantees
the persistence of the past. The idea is closely related to the notion
of ``generalized '' records that we treated in our earlier work \cite{GH90b}.
The physical picture is that, at every branching of the coarse-grained
histories
of the universe, each of the exhaustive and mutually exclusive possibilities
is correlated with a different state of something like a photon or neutrino
going off to infinity and unaffected by subsequent alternatives.
The orthogonality of those states is the realistic mechanism underlying
decoherence. For each of the alternative coarse-grained histories up to
some time,
a projection operator $R$ describes the information of that kind that has
been stored up. The projections
constitute the ``generalized records'' associated with the
different histories. They are all orthogonal to one another
 and that orthogonality gives
rise to the decoherence of histories.

The present work assumes a pure state for the universe, that is, a density
matrix
$\rho$ of the form $|\Psi \rangle \langle \Psi |$. In some earlier articles
\cite{GH90b}
we allowed $\rho$ to be more general. We then defined another kind of ``strong
decoherence'' which was equivalent, for a pure state,  to medium decoherence.
We now suggest restricting the term ``strong decoherence''
to what we are discussing here
and abandoning it as a name for the earlier concept, which may be too
restrictive
when the state is not pure and is redundant otherwise \cite{GH93a}.
For the rest of the article we assume that $\rho =|\Psi\rangle\langle\Psi|$.

The present strong decoherence condition is stronger than our earlier ``medium
decoherence'',
which in turn is stronger than our ``weak decoherence'' condition. Other
authors have
discussed still weaker conditions, for example,
the ``consistent histories'' condition of Griffiths \cite{Gri84}
 and Omn\`es \cite{Omnsum}
and the linearly positive histories of Goldstein and Page \cite{GP95}.
A simple and instructive
case of the present strong decoherence was discussed in an insightful paper by
Finkelstein \cite{Fin93}, who called it ``PT-decoherence'' and showed
how it is related to the
``decoherence of density matrices'' that has been discussed by many
({\it e.g.}~\cite{Zeh71,Zursum,UZ89,PZ93}).  We shall consider this
relationship in a more general context and show how a variety of reduced
density matrices can be constructed for individual histories
up to a given time that are diagonal in appropriate alternatives at the next
time as a consequence of strong decoherence.

In Section II we shall
review the various decoherence conditions after introducing some
necessary notation. Section III introduces strong decoherence and
describes
the connection with reduced density matrices. Section IV explores these
ideas in simplified models in which the coarse grainings are restricted to
those that
follow one set of fundamental co\"ordinates while ignoring all others.
In Section V we review our program to provide a measure of classicality and
discuss the role that strong decoherence might play in such a program.

\section
{Varieties of Decoherence}
\label{sec:II}
The ideas of the quantum mechanics of closed systems, including the (medium)
decoherence of sets of
alternative coarse-grained histories, can be formulated in perfect
generality for quantum field theory. One can include the effects of a quantized
spacetime metric, as
in a field theory (Lagrangian) version of superstring theory, by  using the
principles of
generalized quantum theory \cite{Har91a,Harpp,Ish94}.
However, it is convenient, as well as an excellent
approximation for many accessible coarse grainings, to consider a fixed
spacetime geometry
with well defined timelike directions. In the following brief review of the
quantum
mechanism of closed systems we shall adopt this approximation, using a time
variable $t$
and the associated Hamiltonian $H$.

One way of specifying a set of alternative histories is to give sets of
alternative
projection operators as a sequence of times $t_1<t_2\cdots<t_n$. At each time
$t_k$, we
have a set of Heisenberg picture projection operators
$\{P^k_{\alpha_k\alpha_{k-1}\cdots\alpha_1}(t_k; t_{k-1},\cdots,t_1)\}$ where
$\alpha_k=1,2,3\cdots$ denotes the particular alternative in the set. The
notation is designed to indicate the branch dependence that is
characteristic of useful sets of alternative coarse-grained histories of
the universe \cite{GH93a}. The different alternatives in the set at
time $t_k$ correspond to different $\alpha_k$. However, in a
branch-dependent
set of histories, the set of alternatives at a given time
will depend on previous history. Useful sets of alternative
histories of the universe (such as those constituting a
quasiclassical realm) will be
branch-dependent because the efficacy of physical
mechanisms of decoherence depends
on particular present circumstances and past history. Branch
dependence is indicated explicitly
by the extended subscript $\alpha_k \alpha_{k-1}\cdots\alpha_1$ and
the dependence on previous times $t_{k-1}\cdots t_1$. The projection
operators are mutually exclusive and exhaustive as expressed by the
relations:

\begin{equation}
P^k_{\alpha_k\alpha_{k-1}\cdots\alpha_1}
P^k_{\alpha^\prime_k\alpha_{k-1}\cdots\alpha_1} =
\delta_{\alpha_k\alpha^\prime_k}
P^k_{\alpha_k\alpha_{k-1}\cdots\alpha_1}\ , \quad
\Sigma_{\alpha_k} P^k_{\alpha_k\alpha_{k-1}\cdots\alpha_1}=I\ ,
\label{twoone}
\end{equation}
where, as will often be convenient,  we have suppressed the time labels for
the sake of compactness. The same physical set of alternatives
can be expressed at later times $t>t_k$ by way of the Heisenberg
equations of motion

\begin{equation}
P^k_{\alpha_k\alpha_{k-1}\cdots\alpha_1}(t; t_{k-1}\cdots t_1)=e^{iH(t-t_k)}
P^k_{\alpha_k\alpha_{k-1}\cdots\alpha_1}(t_k; t_{k-1}\cdots t_1)
e^{-iH(t-t_k)}\ .
\label{twotwo}
\end{equation}
(Here, and throughout, we use units such that $\hbar = 1$.)

Each history is then a particular sequence of alternatives
$\alpha=(\alpha_1, \cdots,
\alpha_k)$ and is represented by the corresponding chain of
projections:\footnote{\setlength{\baselineskip}{.1in}These
are called {\sl homogeneous} histories by Isham \cite{Ish94}.
We used a slightly
different notation in \cite{GH93a}
 with $P^k_{\alpha_k}(t_k; \alpha_{k-1},t_{k-1},
\cdots \alpha_1)$ instead of $P^k_{\alpha_k\cdots \alpha_1}(t_k;
t_{k-1}\cdots t_1)$.}
\begin{equation}
H_{\alpha_k\cdots\alpha_1}  =  P^k_{\alpha_k\cdots\alpha_1}(t_k;
t_{k-1}, \cdots, t_1)
P^{k-1}_{\alpha_{k-1}\cdots
\alpha_1}(t_{k-1}; t_{k-2}, \cdots, t_1)\cdots P^1_{\alpha_1}(t_1)\ .
\label{twothree}
\end{equation}
As mentioned above, we shall not
always indicate the various times.
Indeed, since any time label may be altered (preserving the order)
by reexpressing
the corresponding projections in terms of field operators at another time using
the equations
of motion, we shall generally suppress these labels.

Sets of histories consisting of chains of projections like (\ref{twothree})
are not the only sets
potentially assigned probabilities by quantum mechanics. As we
have mentioned, the general notion of coarse-graining is a partition of a
fine-grained set into classes $c_\beta, \beta=1,2,\cdots$. Such histories may
consist of {\it sums} of chains
\begin{equation}
C_\beta = \sum\limits_{(\alpha_1, \cdots, \alpha_k)\in\beta}
H_{\alpha_k\cdots\alpha_1}\ .
\label{twofour}
\end{equation}
Evidently,
\begin{equation}
\sum\limits_\beta C_\beta=I\ .
\label{twofive}
\end{equation}

Various authors have discussed different conditions for when a set of histories
$\{c_\beta\}$ decoheres and can be assigned probabilities $p(\beta)$ in quantum
theory. Below we list them in increasing order of strength --- first for an
initial
condition described by a density matrix $\rho$ and then for the special case
that $\rho$
is pure, $\rho=|\Psi\rangle\langle\Psi|$.

\begin{itemize}
\item The ``linearly positive'' condition of Goldstein and Page \cite{GP95}:
\begin{mathletters}
\label{twosix}
\begin{eqnarray}
p(\alpha)&=&Re\ Tr(C_\alpha\rho)\geq 0\ , \label{twosixa}\\
p(\alpha)&=&Re \langle\Psi|C_\alpha|\Psi\rangle\geq 0\ . \label{twosixb}
\end{eqnarray}
\end{mathletters}
\item The ``consistent histories'' condition of Griffiths \cite{Gri84} and
Omn\`es \cite{Omnsum} for sets of histories that are chains of the form
(\ref{twothree})
(homogeneous sets):
\begin{mathletters}
\label{twoseven}
\begin{eqnarray}
Re\ Tr(C_{\alpha^\prime}\rho C^\dagger_\alpha) & = &
\delta_{\alpha^\prime\alpha}\, p(\alpha)\ , \label{twosevena}\\
Re\ \langle\Psi|C^\dagger_\alpha C_{\alpha^\prime}|\Psi\rangle &=&
\delta_{\alpha^\prime\alpha}\, p(\alpha)\ ,\label{twosevenb}
\end{eqnarray}
\end{mathletters}
{\it provided} $C_\alpha + C_\alpha'$ is a chain as well.
\item Weak decoherence:
\begin{mathletters}
\label{twoeight}
\begin{eqnarray}
Re\ Tr(C_{\alpha^\prime}\rho\, C^\dagger_\alpha) &=&
\delta_{\alpha^\prime\alpha}\, p(\alpha)\ ,\label{twoeighta}\\
Re\ \langle\Psi|C^\dagger_\alpha
C_{\alpha^\prime}|\Psi\rangle &=& \delta_{\alpha^\prime\alpha}
\, p(\alpha)\ ,\label{twoeightb}
\end{eqnarray}
\end{mathletters}
with no restriction to chains or on the sums of $C'$s.
\item Medium decoherence:
\begin{mathletters}
\label{twonine}
\begin{eqnarray}
Tr(C_{\alpha^\prime}\rho\, C^\dagger_\alpha) &=&
\delta_{\alpha^\prime\alpha}\, p(\alpha)\ ,\label{twoninea}\\
\langle\Psi|C^\dagger_\alpha C_{\alpha^\prime}|\Psi\rangle
&=& \delta_{\alpha^\prime\alpha}\, p(\alpha)\ ,\label{twonineb}
\end{eqnarray}
\end{mathletters}
again with no restrictions on the $C'$s.
\end{itemize}

In this paper we shall discuss a yet stronger condition of decoherence, which
for
a pure state has the form
\begin{equation}
\langle\Psi|C^{\dagger}_\alpha M^{\dagger} M' C_{\alpha'}|\Psi\rangle=0, \quad
\alpha \neq \alpha'\ ,
\label{twoten}
\end{equation}
for any operators $M$ included in a set $\{M\}_\alpha$ and $M'$ included in a
set $\{M\}_{\alpha'}$, both sets including the identity, $I$. We shall write
this as
\begin{equation}
\langle\Psi|C^\dagger_\alpha\{M^{\dagger}\}_\alpha
\{M^\prime\}_{\alpha^\prime}C_{\alpha^\prime}
|\Psi\rangle = 0 \quad \alpha^\prime
\not=\alpha\ .
\label{twoeleven}
\end{equation}
where the occurrence of $\{M\}$ in an equation means that it holds for
each $M\in\{M\}$.
The probabilities for a set satisfying this condition are
\begin{equation}
p(\alpha)=\langle\Psi|C^{\dagger}_\alpha C_\alpha |\Psi\rangle\ .
\label{twotwelve}
\end{equation}
The properties of the sets $\{M\}_\alpha$ that make this a condition of strong
decoherence are discussed in the next section.

\section
{Strong Decoherence}
\label{sec:III}

We shall now introduce a notion of strong decoherence applying to a set of
histories that are chains of projections. This strong decoherence is a
special form of medium decoherence and thus permits the assignment of
probabilities to any set of histories that is a coarse graining of the
set of chains, whether or not the sums of chains involved are themselves
chains. Such coarse grainings can also be considered to be strongly
decoherent.

The definition of strong decoherence is connected with the properties of
generalized records of histories that we have described in earlier work.
When a set of histories obeys medium decoherence and the initial
condition is a pure state $|\Psi\rangle$, then the branch states
$C_\alpha|\Psi\rangle$ corresponding to the individual histories are
mutually orthogonal. There is, therefore, a set of orthogonal projection
operators $\{R_\alpha\}$ on the branches. Indeed, unless the branches
constitute a basis in Hilbert space (a full set
of histories), there will be many different possible choices of the
$\{R_\alpha\}$. Thus, medium decoherence implies
\begin{equation}
C_\alpha|\Psi\rangle = R_\alpha|\Psi\rangle
\label{threeone}
\end{equation}
for various sets of projections satisfying
\begin{equation}
R_\alpha R_{\alpha'} =\delta_{\alpha\alpha'}R_\alpha\ .
\label{threetwo}
\end{equation}
The $R_\alpha$ are generalized records of the histories $C_\alpha$.
We call them ``generalized'' records because it does not follow from
this definition alone that these operators have any of the further
properties that might normally be required of records --- being
quasiclassical operators, persisting over a period of time, being
accessible to an observer, etc.

Medium decoherence and a pure initial state  imply the existence of
the generalized
records
as we have just described, but conversely the existence of generalized
records yields medium decoherence. To see this, merely note that
eqs.~(\ref{threeone})
and (\ref{threetwo}) imply the orthogonality of the branches
$C_\alpha|\Psi\rangle$
and that is what medium decoherence means. In this paper we are going to
discuss generalized records of a special kind.

We are going to introduce at any stage in history a pool of operators from
which future history may be drawn so as to ensure the permanence of the
past. This pool will contain some
 operators that are very different from
the record operators, as in the realistic mechanisms we have described, but
correlated with those record operators through the state $|\Psi\rangle$.
In general, the histories will consist of chains of projections that have
this character (or sums of such chains). To that end, it is useful, for
a given choice of generalized records, to consider the set of operators $M$
that commute with the $R_\alpha$ when acting on the branch state vector,
that is
\begin{mathletters}
\label{threethree}
\begin{eqnarray}
[M,R_\alpha]C_\alpha|\Psi\rangle &= &0, \label{threethreea}
\end{eqnarray}
or
\begin{eqnarray}
[M,R_\alpha]R_\alpha|\Psi\rangle &=&0\ .\label{threethreeb}
\end{eqnarray}
\end{mathletters}
We denote this linear space of operators by $\{M\}_\alpha$.
Writing out (\ref{threethree}) and using $(R_\alpha)^2 =R_\alpha$,
one finds the equivalent
condition
\begin{equation}
R_\alpha M R_\alpha|\Psi\rangle=M R_\alpha |\Psi\rangle.
\label{threefour}
\end{equation}
This says that $MR_\alpha|\Psi\rangle $ lies in the subspace designated by
$R_\alpha$. Thus for $M\in\{M\}_\alpha$ and $M'\in\{M\}_{\alpha'}$,
we have that $MR_\alpha|\Psi\rangle$ is orthogonal to $M'R_{\alpha'}|\Psi
\rangle$, or, what is the same thing,
$MC_\alpha|\Psi\rangle$ is orthogonal to $M'C_{\alpha'}|\Psi\rangle$.

The sets $\{M\}_\alpha$ that effectively commute with the generalized
records $R_\alpha$ are thus operators for which
\begin{equation}
\bigl\langle\Psi|C^\dagger_\alpha \{M^{\dagger}\}_\alpha
\{M^\prime\}_{\alpha^\prime} C_{\alpha^\prime}|\Psi\bigr\rangle = 0,
\quad \alpha^\prime\not=\alpha\ ,
\label{threefive}
\end{equation}
as advertised in (\ref{twoeleven}).

Each $R_\alpha$ can be expressed in terms of the operators in $\{M\}_\alpha$
and the branch states $C_\alpha|\Psi\rangle$. In fact,
\begin{equation}
R_\alpha ={\rm Proj}\, (\{M\}_\alpha C_\alpha|\Psi\rangle).
\label{threesix}
\end{equation}
by which we mean that $R_\alpha$ projects onto the subspace spanned by the
vectors of the form $MC_\alpha|\Psi\rangle$ for $M\in\{M\}_\alpha$.
In the future we shall use the notation $R_\alpha$ both for the operator and
the subspace onto which it projects. To derive (\ref{threesix}), one simply has
to note that, if $|v\rangle$ is a vector in $R_\alpha$, then it necessarily has
the form
$MR_\alpha|\Psi\rangle$ for some operator $M$ that effective commutes with
$R_\alpha$ as in (\ref{threethree}). (The operator
$M=|v\rangle\langle\Psi|R_\alpha$ is
a simple example.) Eq.~(\ref{threesix}) then follows.

The existence of
 generalized records of some kind and their concomitant $M$'s follows from
medium decoherence and the assumption of a pure state. The decoherence
condition (\ref{threefive}) is therefore no stronger than medium decoherence
unless
the generalized records or the concomitant $M$'s are restricted in some
way. We shall shortly define strong decoherence by  conditions on the records
that guarantee the permanence of the past. Before doing so, however, it
is useful to spell out how the above discussion would go if the $M$'s were
specified first rather than the $R$'s.

Suppose we are given sets $\{M\}_\alpha$ that satisfy
(\ref{threefive}). We require all the sets $\{M\}_\alpha$
to contain $I$, and so such decoherent sets of histories will be medium
decoherent and generalized records satisfying (\ref{threeone}) will exist.
There
are many choices for these records, but an attractive one is
(\ref{threesix}).
These $R$'s are orthogonal [cf.~(\ref{threetwo})] as a consequence of
(\ref{threefive}). They are
generalized records satisfying (\ref{threeone}) because, with
$M=I$ contained in
$\{M\}_\alpha$, the branch state vector
 $C_\alpha|\Psi\rangle$ is in the subspace $R_\alpha$.
Thus
\begin{equation}
R_\alpha C_\beta |\Psi\rangle = \delta_{\alpha\beta} C_\beta |\Psi\rangle.
\label{threeseven}
\end{equation}
Summing over $\beta$ gives (\ref{threeone}).
It is then an easy computation to show that, as in (\ref{threethree}), the
$M$'s effectively commute with the $R$'s defined by (\ref{threesix}).
Eq.~(\ref{threesix}) is
not the most general identification of records with this commutation
property. One could add to the $\{R_\alpha\}$ any mutually orthogonal
projectors $V_\alpha$ that are also orthogonal to all the
$R_\alpha$ given by (\ref{threesix}). If that is done with
non-vanishing $V_\alpha$,
then the class of operators effectively commuting  with $R_\alpha$ is    wider
than the $\{M\}_\alpha$, containing also all operators leading from
$C_\alpha |\Psi\rangle$ to the space $V_\alpha$.

{From} now on, we shall adopt the point of view that the $V$'s are zero
and (\ref{threesix}) holds. That way either the $R$'s or the $M$'s could
be regarded as primary. We shall also
 deal with explicitly decoherent sets of histories
composed of chains of projectors, bearing in mind that we could also treat
any coarse grainings of such sets.

We are now in a position to define a notion of strong decoherence that
will provide a systematic, step by step procedure for extending sets of
histories into the future so as to guarantee the permanence of the past.
We restrict attention to sets of homogeneous histories represented by
branch-dependent chains of the form (\ref{twothree}) at $n$ times
$t_1,\cdots, t_n$. Partial histories up to time $t_\ell$ are represented by
chains
\begin{equation}
H^\ell_{\alpha_\ell\cdots\alpha_1}= P^\ell_{\alpha_\ell\cdots\alpha_1}
P^{\ell-1}_{\alpha_{\ell-1}\cdots\alpha_1}\cdots P^1_{\alpha_1}\ ,
\label{threeeight}
\end{equation}
where we have introduced a superscript $\ell$ on
$H^\ell_{\alpha_\ell\cdots\alpha_1}$ to indicate that it refers to the first
$\ell$ times. The definition of a strongly decoherent set of histories
concerns the relation between the histories at a time $t_\ell$ and those
for all future times up to $t_n$, for each $t_\ell$.

Strong decoherence is a special case of medium decoherence, so that all
the sets $\{H^\ell_{\alpha_\ell\cdots\alpha_1}\}\ , \ell =1, \cdots, n$
satisfy the condition (\ref{twonine}). For each $\ell$ there is a set of
generalized
records $\{R^\ell_{\alpha_\ell\cdots\alpha_1}\}$. From these, the sets of
operators $\{M\}^\ell_{\alpha_\ell\cdots\alpha_1}$ that effectively
commute with the $R^\ell_{\alpha_\ell\cdots\alpha_1}$
[cf.~(\ref{threethree})] may
be constructed.

Two conditions define a strongly decoherent set. The first is that at
each $\ell$ and for each history $\alpha_\ell\cdots\alpha_1$, the set of
operators $\{M\}^\ell_{\alpha_\ell\cdots\alpha_1}$, suitably adjusted
for evolution, forms the pool from which future histories are drawn.
Specifically, we assume for each $\ell < k \leq n$

\begin{equation}
{\rm I.} \qquad\qquad\qquad\qquad
P^k_{\alpha_k\cdots\alpha_\ell\cdots\alpha_1} \cdots
P^{\ell+1}_{\alpha_{\ell+1}\alpha_\ell\cdots\alpha_1} \in
U_{k\ell}\{M\}^\ell_{\alpha_\ell\cdots\alpha_1}\ ,
\label{threenine}
\end{equation}
where $U_{k\ell}$ is a unitary operator that adjusts the $M$'s in the
set $\{M\}^\ell_{\alpha_\ell\cdots\alpha_1}$ to those appropriate to a later
time. The range of possibilities for the $U_{k\ell}$, as well as the
need for this kind of adjustment, will become clearer when we discuss
particular models in the next section. We note that if histories are
extended by choosing the future alternatives from among the present
$M$'s by (\ref{threenine}), the extended set will continue to be medium
decoherent. That follows from (\ref{threefive}) and the unitary character
of the $U$'s.  The permanence
of the past is thus guaranteed in these extensions to the future. In
particular, if we  construct strongly decoherent sets step by step
by choosing the $P$'s at the next time from the pool of $M$'s at this
time, the permanence of the past is assured.

Note that the condition (\ref{threenine}) incorporates branch
dependence. The pool of operators for extending each history depends on
that history. Note also, that the condition (\ref{threenine}) is already
more restrictive than just medium decoherence.  We showed that the
operators that effectively commute with $R^k_{\alpha_k\cdots\alpha_1}$
satisfy (\ref{threefive}), but we did not show that every possible
medium decoherent extension commutes with the
$R^k_{\alpha_k\cdots\alpha_1}$.

To enforce the condition (\ref{threenine}) for each pair of times $\ell$
and $k$ such that $\ell < k \leq n$ requires a consistency condition
between the sets of $M$'s at different times, relations between the
adjustment operators $U$,  and restrictions on the
concomitant records. The consistency condition for the $U$'s is the the
elementary requirement of composition:
\begin{equation}
U_{k\ell}U_{\ell j}=U_{kj}.  \  \label{harry}
\end{equation}
To discuss the requirements on the $M$'s, it is useful to introduce the
following notation for $\ell < k \leq n$:
\begin{eqnarray}
\{M\}^k_{\alpha_\ell\cdots\alpha_1} & \equiv & U_{k\ell}
\{M\}^\ell_{\alpha_\ell\cdots\alpha_1}\ , \\
\label{threeeleven}
\{R\}^k_{\alpha_\ell\cdots\alpha_1} & \equiv & U_{k\ell}
\{R\}^\ell_{\alpha_\ell\cdots\alpha_1} U^\dagger_{k\ell}\ .
\label{threetwelve}
\end{eqnarray}
The defining condition (\ref{threenine}) may then be restated
\begin{eqnarray}
P^k_{\alpha_k\cdots\alpha_\ell\cdots\alpha_1} \cdots
P^{\ell+1}_{\alpha_{\ell+1}\alpha_\ell\cdots\alpha_1} \in
\{M\}^k_{\alpha_\ell\cdots\alpha_1}\ .
\label{threethirteen}
\end{eqnarray}
Consistency requires that not only the future histories be drawn from
the pool of $M$'s, but also the future $M$'s as well.
At the weakest this requires that
\begin{equation}
{\rm Proj}\,\left[\{M\}^k_{\alpha_k\cdots\alpha_\ell\cdots\alpha_1}
P^k_{\alpha_k\cdots\alpha_\ell\cdots\alpha_1}
\cdots P^{\ell+1}_{\alpha_{\ell+1}\cdots\alpha_1}
H^\ell_{\alpha_\ell\cdots\alpha_1}|\Psi\bigr\rangle\right]
\subseteq
{\rm Proj}\,\left[\{M\}^k_{\alpha_\ell\cdots\alpha_1}
H^\ell_{\alpha_\ell\cdots\alpha_1}|\Psi\bigr\rangle\right]\ .
\label{threefourteen}
\end{equation}

According to (\ref{threesix}) the record operators
$R^k_{\alpha_k\cdots\alpha_1}$ are the projections onto the subspaces
spanned by the vectors $\{M\}^k_{\alpha_k\cdots\alpha_1}
H^k_{\alpha_k\cdots\alpha_1}|\Psi\rangle$. A consequence of the
consistency condition (\ref{threethirteen}) is therefore
\begin{equation}
R^k_{\alpha_k\cdots\alpha_1} \subseteq R^k_{\alpha_\ell\cdots\alpha_1}
\ ,\ \ell < k \leq  n\ .
\label{threefifteen}
\end{equation}
Thus the records nest. Eq.~(\ref{threefifteen}) means that, up to
unitary adjustment, the subspaces corresponding to the records narrow
with each extension into the future. In this sense the records of the
earlier parts of history are preserved in records of the later parts.

Nesting of records (\ref{threefifteen}) implies the operator form of
the consistency relation (\ref{threefourteen}). To see this suppose that
$M\in \{M\}^k_{\alpha_k\cdots\alpha_1}$. This means that $M$ effectively
commutes with $R^k_{\alpha_k\cdots\alpha_1}$, or explicitly
[cf.~(\ref{threefour})]
\begin{equation}
R^k_{\alpha_k\cdots\alpha_1} M\, H^k_{\alpha_k\cdots\alpha_1}
|\Psi\bigr\rangle = M\, H^k_{\alpha_k\cdots\alpha_1} |\Psi\bigr\rangle
\ .
\label{threesixteen}
\end{equation}
This can be rewritten
\begin{equation}
R^k_{\alpha_k\cdots\alpha_1} \left(M\,
P^k_{\alpha_k\cdots\alpha_1}\cdots
P^{\ell+1}_{\alpha_{\ell+1}\cdots\alpha_1}\right)\,
H^\ell_{\alpha_\ell\cdots\alpha_1} |\Psi\bigr\rangle
= \left(M\, P^k_{\alpha_k\cdots\alpha_1} \cdots
P^{\ell+1}_{\alpha_{\ell+1}\cdots\alpha_1}\right)
\, H^\ell_{\alpha_\ell\cdots\alpha_1} |\Psi\bigr\rangle\ .
\label{threeseventeen}
\end{equation}
But, since $R^k_{\alpha_k\cdots\alpha_1}$ is contained within
$R^k_{\alpha_\ell\cdots\alpha_1}$, this means that $M\, P^k_{\alpha_k}\cdots
P^{\ell+1}_{\alpha_{\ell+1}}$ must effectively commute with
$R^\ell_{\alpha_k\cdots\alpha_1}$. Thus, $M\, P^k_{\alpha_k\cdots}\cdots
P^{\ell+1}_{\alpha_{\ell+1}}\in \{M\}^k_{\alpha_\ell\cdots\alpha_1}$.
This is the operator form of the consistency relation (\ref{threefourteen})
\begin{equation}
\{M\}^k_{\alpha_k\cdots\alpha_1} P^k_{\alpha_k\cdots\alpha_1}\cdots
P^{\ell+1}_{\alpha_{\ell+1}\cdots\alpha_1} \in
\{M\}^k_{\alpha_\ell\cdots\alpha_1}\ .
\label{threeeighteen}
\end{equation}
Effective consistency (\ref{threefourteen})
thus implies the necessary operator form
(\ref{threeeighteen}).

The nesting of generalized records (\ref{threefifteen}) can be
regarded as a modern form of a very old idea.  If the generalized
records were to consist of genuinely independent ``registrations'' in
commuting variables (``memory slots''), then we would have
\begin{equation}
R^k_{\alpha_k\cdots\alpha_1} = \prod\limits^k_{j=1}\ A^{(j)}_{\alpha_j}
\ ,
\label{threenineteen}
\end{equation}
where the $\{A^{(j)}_{\alpha_j}\}$ are a set of orthogonal projection
operators for each $j$, all the operators in each set commuting with all
those in other sets with
different $j$. In that case proceeding to the next time
would result in the registration of a new record in a further
exhaustive set of alternatives $\{A^{(k+1)}_{\alpha_{k+1}}\}$ commuting
with all the previous ones. The nesting condition (\ref{threefifteen})
would then follow immediately. In the jargon of the early exponents of
 quantum mechanics, the
$A$'s, and hence the $R$'s, could be regarded as ``$c$-numbers''.
The non-commuting operators $P^k_{\alpha_k\cdots\alpha_1}\cdots
P^1_{\alpha_1}$ (``$q$-numbers'') are registered in commuting
``$c$-numbers'' just as in the old theory of
measurement\footnote{\setlength{\baselineskip}{.1in}By
means of generalized records the universe could be said to be measuring
itself. However, we do not recommend this terminology because, as we
have stressed before, these records need not be quasiclassical operators
or accessible to an observer as they were in some formulations of the
old theory of measurement.} (See, {\it e.g.} \cite{Wig63}).

The nesting inclusion relation (\ref{threefifteen}) implies
\begin{equation}
\sum\limits_{\alpha_k\cdots\alpha_{\ell+1}}
R^k_{\alpha_k\cdots\alpha_\ell\cdots\alpha_1} \subseteq
R^k_{\alpha_\ell\cdots\alpha_1}\ ,
\label{threetwenty}
\end{equation}
since the $R$'s are a set of orthogonal projection operators. It is
implausible that in realistic situations this inclusion relation becomes
an equality. Equality in the simplified model just described would
require that the $\{A^{(j)}_{\alpha_j}\}$ be an exhaustive set of projections
summing to unity. But there is no particular reason that the alternative
configurations of the ``memory slots'' that record history should
exhaust all possible configurations.

However, {\it effective}
 equality {\it is} plausible. That is because it follows, by
summing both sides of (\ref{threeone}), that
\begin{equation}
\sum\limits_{\alpha_k\cdots\alpha_{\ell+1}}
R^k_{\alpha_k\cdots\alpha_\ell\cdots\alpha_1}|\Psi\bigr\rangle =
R^\ell_{\alpha_\ell\cdots\alpha_1}|\Psi\bigr\rangle.
\label{threetwentyone}
\end{equation}
The effective equality corresponding to the inclusion relation
 (\ref{threetwenty}) would thus mean
\begin{equation}
{\rm II.}\qquad\qquad\qquad R^\ell_{\alpha_\ell\cdots\alpha_1}
|\Psi\bigr\rangle \approx
R^k_{\alpha_\ell\cdots\alpha_1}|\Psi\bigr\rangle\ , k > \ell\ .
\label{threetwentytwo}
\end{equation}
This relation expresses the permanence of the record of the history
$(\alpha_1,\cdots,\alpha_\ell$) that we expect in physically realistic
mechanisms of decoherence.
We take it to be the second defining property of strong decoherence.

Eq.~(\ref{threetwentytwo}) cannot be exactly
satisfied because it holds only for times $t_k$ later than $t_\ell$.
The vanishing of the
difference between left and right hand sides of (\ref{threetwentytwo}) just
for $t_k>t_\ell$ cannot
be strictly true in quantum mechanics, where evolution is continuous and
amplitudes analytic, but it can be true to an excellent approximation,
which improves rapidly as the interval
increases.\footnote{\setlength{\baselineskip}{.1in}We have
previously discussed how the creation of permanent
records of alternative histories for subsystems
can produce decoherence in connection
with ideal measurement models. (See \cite{Har91a}, Section II.10). These models
show how the usual ``Copenhagen'' quantum mechanics of measured subsystems
is an approximation to the more general quantum mechanics of closed
systems under discussion here. To the extent that these idealized models
reflect typical experimental situations, strong decoherence may be
said to be assumed in the ``Copenhagen'' approximation.}
Indeed, physically one cannot  expect it to hold until a time after
$t_\ell$ when the record is created.

Many authors have investigated the properties of reduced density matrices
in connection with models of the mechanisms of decoherence
({\it e.g.}~\cite{Zeh71,Zursum,UZ89,PZ93}).
In the framework we have been discussing, a variety of
``reduced'' density matrices  can be constructed
for each history up to a given time. They  are  generalizations of the kinds
considered in those works.
Strong decoherence guarantees that
these  ``reduced'' density matrices are diagonal with respect to the
alternatives at the next time in the history.  A very
useful example of such a construction is as follows:

To be definite, fix attention
on the history
$(\alpha_1,\alpha_2)$ and consider the sets
$\{M\}^3_{\alpha_3\alpha_2\alpha_1}$ for the various possible values of
$\alpha_3$. ( These are  the sets of operators from which the alternatives
in the chains of
histories at time $t_3$ are drawn.) Take any {\it common} subset of all these
sets that is
closed under Hermitian conjugation and multiplication as well as addition and
multiplication by complex numbers.   More succinctly, take a common subset
that is an {\it algebra} closed under Hermitian conjugation. Denote  this
algebra by $\{F\}^3_{\alpha_2\alpha_1}$ and let $\{A^{\alpha_2\alpha_1}_i\}$
be a Hermitian operator basis for it conveniently normalized by
$Tr(A^{\alpha_2\alpha_1}_i A^{\alpha_2\alpha_1}_j)=\delta_{ij}$.
A reduced density matrix may be specified by applying
the Jaynes construction to the coarse-graining defined by the operators
$\{F\}^3_{\alpha_2\alpha_1}$. That is, we define $\tilde\rho^{\alpha_2
\alpha_1}$ as the density matrix that maximizes the entropy functional
$-Tr(\tilde\rho \log \tilde\rho)$ subject to the constraint
of reproducing  the correct expectation values of the $F$'s. Specifically,
$\tilde\rho^{\alpha_2\alpha_1}$ maximizes entropy subject to the constraints
\begin{equation}
Tr(A^{\alpha_2\alpha_1}_i \tilde\rho) = \langle \Psi^2_{\alpha_2\alpha_1}
|A^{\alpha_2\alpha_1}_i |\Psi^2_{\alpha_2\alpha_1}\rangle \ .  \label{moe}
\end{equation}
Here $|\Psi^2_{\alpha_2\alpha_1}\rangle$ is the normalized branch state vector
corresponding to the history $(\alpha_1,\alpha_2)$:
\begin{equation}
|\Psi^2_{\alpha_2\alpha_1}\rangle \propto H^2_{\alpha_2\alpha_1}|\Psi\rangle
=R^2_{\alpha_2\alpha_1}|\Psi\rangle \ . \label{henry}
\end{equation}
Since the set $\{F\}^3_{\alpha_2\alpha_1}$ is closed under
multiplication, $\tilde\rho^{\alpha_2\alpha_1}$ may be computed explicitly
(see, {\it e.g} \cite{GH96}), and comes out
\begin{equation}
\tilde\rho^{\alpha_2\alpha_1}= \sum\limits_i
\langle\Psi_{\alpha_2\alpha_1}|A^{\alpha_2\alpha_1}_i
|\Psi_{\alpha_2\alpha_1}\rangle A^{\alpha_2\alpha_1}_i \ . \label{ed}
\end{equation}

Strong decoherence implies that the density matrix defined by eq.
(\ref{ed}) is diagonal in the next alternatives in the chain ---
those labeled by $\alpha_3$. To see this
note that because $\{F\}^3_{\alpha_2\alpha_1}$ is a subset of
$\{M\}^3_{\alpha_3\alpha_2\alpha_1}$ for any $\alpha_3$,
\begin{equation}
\langle\Psi^3_{\alpha'_3\alpha_2\alpha_1}|\{F\}^3_{\alpha_2\alpha_1}|
\Psi^3_{\alpha_3\alpha_2\alpha_1}\rangle =0, \quad \alpha'_3 \ne
\alpha_3 \ . \label{tom}
\end{equation}
The $\{A^{\alpha_2\alpha_1}_i\}$ are contained in the
$\{F\}^3_{\alpha_2\alpha_1}$, so that
\begin{equation}
\langle\Psi^3_{\alpha'_3\alpha_2\alpha_1}|\tilde\rho^{\alpha_2\alpha_1}|
\Psi^3_{\alpha_3\alpha_2\alpha_1}\rangle =0, \quad \alpha'_3 \ne
\alpha_3 \ . \label{tom2}
\end{equation}
Thus, the reduced density matrix defined
by (\ref{ed}) is diagonal in $\alpha_3$.
We shall make the connection  with the
usual construction of reduced density matrices more concrete
in the context of the the model discussed in the next Section.

The important observation in the above argument is the vanishing of the matrix
elements (\ref{tom}).
We could replace the whole notion of
a reduced density matrix by eq.(\ref{tom}). Indeed,
because $\{F\}^3_{\alpha_2\alpha_1}$ is closed under
multiplication, a much stronger relation holds
\begin{equation}
\langle\Psi^3_{\alpha'_3\alpha_2\alpha_1}|\{F\}^3_{\alpha_2\alpha_1}
\{F\}^3_{\alpha_2\alpha_1}\{F\}^3_{\alpha_2\alpha_1}
|\Psi^3_{\alpha_3\alpha_2\alpha_1}\rangle =0, \quad \alpha'_3 \ne
\alpha_3 \ . \label{tom1}
\end{equation}
Thus, for any two different values of $\alpha_3$,
the matrix elements of the $F$'s do not simply vanish
between two branch states
$|\Psi^3_{\alpha'_3\alpha_2\alpha_1}\rangle$,
and
$|\Psi^3_{\alpha_3\alpha_2\alpha_1}\rangle$
for distinct $\alpha'_3$ and $\alpha_3$,
but between any vectors in the two subspaces
$\{F\}^3_{\alpha_2\alpha_1}|\Psi^3_{\alpha'_3\alpha_2\alpha_1}\rangle$
and
$\{F\}^3_{\alpha_2\alpha_1}|\Psi^3_{\alpha_3\alpha_2\alpha_1}\rangle$.
Such sets of operators $\{F\}^3_{\alpha_3\alpha_2}$ are not difficult to
find. One example is the set of operators commuting with the
$R^3_{\alpha_3\alpha_2\alpha_1}$ for all values of $\alpha_3$.
As the $R$'s in physically interesting examples are typically very large
subspaces,  there are many subalgebras that
also give possible sets of $F$'s. We will illustrate
all this in the context of a concrete model in the next Section.

The closure of the set of $F$'s under Hermitian conjugation and
multiplication gives the relation (\ref{tom1}), an elegant connection
with the  method of Jaynes,
and, as we shall see, a close connection with
the models frequently discussed. But these assumptions are not necessary
if all we want is to define a reduced
density matrix for the history $(\alpha_1,\alpha_2)$. Consider, for
instance, products
of the form $M'^{\dagger}M$ for $M\in\{M\}^3_{\alpha_3\alpha_2\alpha_1}$ and
$M'\in\{M\}^3_{\alpha'_3\alpha_2\alpha_1}$ for various values of
$\alpha'_3$ and $\alpha_3$.
Suppose all these sets have in common a
closed linear set of operators (depending on $(\alpha_1,\alpha_2)$)
but not necessarily closed under Hermitian conjugation and
multiplication. There is still an operator basis and, if we denote
that basis by
$\{A^{\alpha_2\alpha_1}_i\}$, the construction (\ref{ed}) still
yields a reduced density matrix and it is still diagonal in $\alpha_3$.
The construction with the $F$'s is a special case of this.
Other examples of ``reduced'' density matrices could be defined by using
different linear subsets  of the set of $M'^{\dagger}M$'s.

Strong decoherence is an idealization and generalization of the
ideas emerging from simple models that posit coarse grainings
following one set of fundamental co\"ordinates (the ``system'') while
ignoring the ``environment'' consisting of the remaining co\"ordinates.
In those models
the decoherence of histories of alternatives of the ``system'' is
effected by the creation of approximately correlated, approximately permanent
records in the
``environment''. The generality afforded by the notion of strong
decoherence
is important because the coarse grainings characterizing the
realistic quasiclassical domain (using values of hydrodynamic variables,
for instance) do not correspond to a distinction between two
kinds of co\"ordinates. In the
following, however, we shall make more explicit contact with these
earlier ideas by investigating
simple models.

\section{A Simplified Model}
\label{sec:IV}

As we mentioned in the last Section, strong decoherence is a
generalization and idealization of the idea that a physical mechanism
for decoherence is the correlation of the strings of non-commuting
operators constituting alternative histories with
commuting operators that are records of those histories. In this Section
we shall make this concrete and explicit in a simple model.
The model is a specialization of a class that we have already discussed
(\cite{GH93a}, Section IV) and we shall rely on that discussion here.

The kind of model we have in mind divides the fundamental co\"ordinates
into two groups, the $x$'s and $Q$'s, with a corresponding factorization
of the Hilbert space ${\cal H} = {\cal H}^x \otimes {\cal H}^{Q^1}\otimes
{\cal H}^{Q^2} \otimes \cdots $. We
assume that the total Hamiltonian describing this system can be written
\begin{equation}
H = H_x (x,p) + H_0 (Q,P) + H_I (x,Q)\ ,
\label{fourone}
\end{equation}
where $H_x$ acts on the $x$'s alone, $H_0$ is a sum of terms each acting
on independently on the Hilbert space ${\cal H}^{Q^i}$ of a single
co\"ordinate $Q^i$, and
$H_I$ is the interaction between all the co\"ordinates.

The coarse grainings studied consist of chains of projections at times
$t_1, t_2,\cdots$ that in the Schr\"odinger picture act only on ${\cal
H}^x$ and represent partitions of some complete set of states in ${\cal
H}^x$ at each time. We denote the wave functions of these complete
orthogonal sets by $\{\phi^1_{r_1} (x)\}$ at time $t_1$, $\{\phi^2_{r_2}
(x)\}$ at time $t_2$, etc., later augmenting this notation to indicate
the branch dependence of the possible sets. If, for instance, the set
$\{\phi^1_{r_1} (x)\}$ is partitioned into classes $\alpha_1,
\alpha_2,\cdots,$ the Schr\"odinger picture projection operator onto the
class $\alpha_k$ is
\begin{equation}
\bigl\langle x^{\prime\prime}\bigl| \hat P^1_{\alpha_k} \bigr| x^\prime
\bigr\rangle = \sum\limits_{r_1\in\alpha_k} \phi^1_{r_1}
(x^{\prime\prime}) \phi^{1*}_{r_1} (x^\prime)
\label{fourtwo}
\end{equation}
We use a hat to distinguish Schr\"odinger picture operators from their
Heisenberg counterparts. The coarse grainings therefore refer only to
alternatives in ${\cal H}^x$ which, in the usual terminology, is
the ``system'' while the Hilbert space of the $Q$'s is the
``environment''.

The basic assumptions of the model are most easily stated in the
Schr\"odinger picture.  We assume a pure wave function $\Psi (x,
Q, t_0)$ at the initial time
and prescribe its evolution to later times. The idea is
that alternatives in $x$ become correlated with record slots in
successive co\"ordinates $Q^1, Q^2, \cdots$ at times $t_1, t_2,
\cdots$. Specifically, we assume that, at these times the wave function
has the form
\begin{mathletters}
\label{fourthree}
\begin{eqnarray}
\Psi (x,Q,t_1)& = &\sum\limits_{\alpha_1} \chi^{1\alpha_1} (Q^1)
\sum\limits_{r_1\in\alpha_1} \phi^1_{r_1} (x) \tilde\chi^1_{r_1} \left(Q^2,
Q^3\cdots\right)
\label{fourthreea}\\
\Psi(x,Q,t_2) &=& \sum\limits_{\alpha_2\alpha_1} \chi^{2\alpha_1}
(Q^1) \chi^{2\alpha_2\alpha_1} (Q^2) \sum\limits_{r_2\in\alpha_2}
\phi^{2\alpha_1}_{r_2} (x) \tilde\chi^{2\alpha_1}_{r_2}
\left(Q^3,Q^4,\cdots\right)
\label{fourthreeb}\\
\Psi(x,Q,t_3) &=& \sum\limits_{\alpha_3\alpha_2\alpha_1}
\chi^{3\alpha_1} (Q^1) \chi^{3\alpha_2\alpha_1} (Q^2)
\chi^{3\alpha_3\alpha_2\alpha_1} (Q^3) \sum\limits_{r_3\in\alpha_3}
\phi^{3\alpha_2\alpha_1}_{r_3} (x) \tilde\chi^{3\alpha_2\alpha_1}_{r_3}
\left(Q^4, Q^5,\cdots\right)\ ,
\label{fourthreec}
\end{eqnarray}
\end{mathletters}
etc.~(We are endeavoring here and elsewhere
 to avoid a {\it d\'ebauche} {\it d'indices} by
quoting specific forms rather than the general case, the form
of which should be immediately apparent).  Here the
$\chi^{k\alpha_j\cdots\alpha_1}(Q^j)$ are orthogonal
functions of a the single co\"ordinate $Q^j$ for time $k$, that is
\begin{equation}
\left(\chi^{k\alpha^\prime_j\alpha_{j-1}\cdots\alpha_1}\ ,\
\chi^{k\alpha_j\alpha_{j-1}\cdots\alpha_1}\right)_{{\cal H}^j} =
\delta^{\alpha^\prime_j\alpha_j}.
\label{fourfour}
\end{equation}
The evolution described above is assumed to hold for each branch. That
is, the individual terms in the sum over $\alpha_1$ in
(\ref{fourthreea}) evolve into the corresponding individual terms in the
sums over $\alpha_1$, in (\ref{fourthreeb}) and (\ref{fourthreec}) and
similarly for the sum over $\alpha_2$ in (\ref{fourthreeb}).

The orthogonal functions $\chi^{k\alpha_3\alpha_2\alpha_1}(Q^3)$
represent the memory slots in which the alternatives represented by the
collection $\{\phi^{3\alpha_2\alpha_1}_{r_3}\}$, $r_3\in \alpha_3$ are
stored at times $k\geq 3$. The central assumption of the model is that
once this registration is accomplished the variable $Q^3$ effectively no
longer interacts with the rest of the system. Specifically, we assume
that as factors of the complete wave function (\ref{fourthree}),
 the $\chi^{k\alpha_j\cdots}$ evolve independently after the time of
registration, {\it viz.}
\begin{equation}
\chi^{k\alpha_j\cdots\alpha_1} (Q^j) = e^{-iH_0(t_k-t_\ell)}
\chi^{\ell\alpha_j\cdots\alpha_1}(Q^j),\quad k\geq\ell\ .
\label{fourfive}
\end{equation}
Were $H_0$ effectively zero when acting on these states, we could say
that the alternatives $\alpha_j$ were registered in unchanging marks.
However, more generally, the marks themselves will evolve according to the
Hamiltonian $H_0$.

This model is close in spirit to that used by Finkelstein \cite{Fin93}
in his discussion of ``$PT$-decoherence''. Finkelstein also used special
coarse grainings that distinguished ``system'' from ``environment''.
The assumption (\ref{fourfive}) for the evolution of the records
accomplishes the same purpose as his more drastic assumption that the
interaction
Hamiltonian vanishes between co\"ordinate $Q^j$ and the rest for times
$t_k>t_j$.

The alert reader will not have failed to see that the above model is a
souped-up version of the usual ideal measurement model expressed in the
language of the quantum mechanics of closed
systems.\footnote{\setlength{\baselineskip}{.1in}See, e.g.,
\cite{Wig63}, or almost any current text in quantum mechanics. For an
exposition from the point of view of the quantum mechanics of closed
systems see (\cite{Har91a}, Section II.10).} The co\"ordinates $x$
are those of the ``measured subsystem'' while the co\"ordinates $Q$
include the apparatus. Our model is therefore subject to all the caveats
associated with the ideal measurement model. In particular, as was shown
by Wigner \cite{Wig52} and Araki and Yanase \cite{AY60}, there are no
realistic Hamiltonians $H$ that will effect the evolution
(\ref{fourthree})
{\it exactly}.\footnote{\setlength{\baselineskip}{.1in}See \cite{Har93b} for a
discussion of this and other limitations in the context of the quantum
mechanics of closed systems.}  It also follows from analyticity that a
relation (\ref{fourfive}) cannot hold exactly after one time and
not before it. The evolution (\ref{fourthree}) and
(\ref{fourfive}) must therefore be understood as holding approximately.

The history operators corresponding to alternatives defined by the
successive collections of $\phi$'s are, in the Schr\"odinger picture,
\begin{equation}
\hat H^{\ell}_{\alpha_\ell\cdots\alpha_1} = \hat P_{\alpha_\ell\cdots\alpha_1}
e^{-iH(t_\ell - t_{\ell-1})} \hat P_{\alpha_{\ell-1}\cdots\alpha_1}
\cdots e^{-iH(t_2-t_1)} \hat P_{\alpha_1},
\label{foursix}
\end{equation}
where we have augmented the notation of (\ref{fourtwo}) to include the
branch dependence of the $\phi$'s. Acting with $\hat
H_{\alpha_\ell\cdots\alpha_1}$ on $\Psi(x, Q, t_0)$ just gives the
branch wave function for the history $(\alpha_\ell, \cdots, \alpha_1)$
at time $t_\ell$,
 {\it e.g.}
\begin{eqnarray}
\hat H^{2}_{\alpha_2\alpha_1} \Psi (x,Q, t_0) &=& \Psi_{\alpha_2\alpha_1}
(x,Q,t_2)\nonumber\\
                                         &\equiv & \chi^{2\alpha_1}
(Q^1)\chi^{2\alpha_2\alpha_1}(Q^2) \sum\limits_{r_2\in\alpha_2}
\phi^{2\alpha_1}_{r_2} (x) \tilde\chi^{\alpha_1}_{r_2} (Q^3, \cdots)
\label{fourseven}
\end{eqnarray}
and similarly for other chains.

{From} the evolution prescribed in (\ref{fourthree}) it is easy to
identify candidate record operators of these histories as projections on
the $\chi$ functions that are correlated with them. Specifically,
\begin{mathletters}
\label{foureight}
\begin{eqnarray}
\hat R^1_{\alpha_1} & = & {\rm Proj}\ \left(\chi^{1\alpha_1}
(Q^1)\right),\label{foureighta}\\
\hat R^2_{\alpha_2\alpha_1} & = & {\rm Proj}\ \left(\chi^{2\alpha_1}
(Q^1) \chi^{2\alpha_2\alpha_1} (Q^2)\right), \label{foureightb}\\
\hat R^3_{\alpha_3\alpha_2\alpha_1} & = & {\rm Proj}\
\left(\chi^{3\alpha_1} (Q^1) \chi^{3\alpha_2\alpha_1}(Q^2)
\chi^{3\alpha_3\alpha_2\alpha_1} (Q^3)\right)\ ,
\label{foureightc}
\end{eqnarray}
\end{mathletters}
etc.~Evidently we have
\begin{equation}
\hat H^{\ell}_{\alpha_\ell\cdots\alpha_1} \Psi (x,Q,t_0) = \hat
R^\ell_{\alpha_\ell\cdots\alpha_1} \Psi (x,Q,t_\ell)\ ,
\label{fournine}
\end{equation}
which is the Schr\"odinger picture representative of the relation
(\ref{threeone}) defining records.

Taking the evolution of the individual $\chi$'s given by
(\ref{fourfive}) into account, we can define
\begin{equation}
\hat R^k_{\alpha_\ell\cdots\alpha_1} = \hat U_{k\ell} \hat
R^\ell_{\alpha_\ell\cdots\alpha_1} \hat U^\dagger_{k\ell},
\label{fourten}
\end{equation}
where
\begin{equation}
\hat U_{k\ell} = \exp\left[-iH_0(t_k-t_\ell)\right].
\label{foureleven}
\end{equation}
Because of the standard relation between Schr\"odinger and
Heisenberg pictures
\begin{equation}
O = e^{iHt} \hat{ O}\, e^{-iHt}\ ,
\label{fourtwelve}
\end{equation}
eq.~(\ref{fourten}) is immediately seen to be the Schr\"odinger picture
representative of the definition (\ref{threeeleven}) with
\begin{equation}
U_{k\ell} = e^{iHt_k} e^{-iH_0(t_k-t_\ell)} e^{-iHt_\ell}.
\label{fourthirteen}
\end{equation}
Thus we identify the $U_{k\ell}$ entering
into the definition of strong decoherence for this choice of records.

The following nesting is immediate from the definition
(\ref{foureight})
\begin{equation}
\hat R^k_{\alpha_k\cdots\alpha_1} \subseteq \hat
R^k_{\alpha_\ell\cdots\alpha_1}\ , \quad \ell < k\ .
\label{fourfourteen}
\end{equation}
{From} (\ref{fourtwelve}) and (\ref{fourthirteen}) this is easily seen to
be the Schr\"odinger picture representative of the general nesting
relation (\ref{threefifteen}). From (\ref{foureight}) we also see that the
nesting relation is not generally an equality because the functions
$\chi^{k\alpha_j\cdots\alpha_1} (Q^j) , \alpha_j = 1, 2, \cdots$ are not
generally a complete basis in ${\cal H}^j$.

However, the equality in (\ref{fourfourteen}) can be seen to be satisfied
effectively, {\it i.e.} when the operator relation is acting on the state. That
is a
simple consequence of the fact that the evolution prescribed in
(\ref{fourthree}) was assumed to hold for each branch. Therefore
projecting on a branch $(\alpha_\ell, \cdots, \alpha_1)$ at time
$t_\ell$ and evolving to time $t_k$ is the same as evolving to time
$t_k$ and projecting on the branch.  At time $t_k$ the projection on the
branch $(\alpha_\ell,\cdots,\alpha_1)$ is the projection on the evolved
functions $\chi$, {\it e.g.}
\begin{mathletters}
\label{fourfifteen}
\begin{eqnarray}
\hat R^k_{\alpha_2\alpha_1} & \equiv & e^{-iH_0(t_k-t_2)}
\hat R^2_{\alpha_2\alpha_1} e^{iH_0(t_k-t_2)}\label{fourfifteena}\\
                            & = & {\rm Proj} \left(\chi^{k\alpha_1}
(Q^1) \chi^{k\alpha_2\alpha_1} (Q^2)\right).\label{fourfifteenb}
\end{eqnarray}
\end{mathletters}
Specifically, therefore, from the form of (\ref{foureight}) and
(\ref{fourten}) we have
\begin{equation}
{\rm II.}\qquad\qquad \hat R^k_{\alpha_\ell\cdots\alpha_1} \Psi (x,Q,t_k) =
e^{-iH(t_k-t_\ell)} \hat R^\ell_{\alpha_\ell\cdots\alpha_1} \Psi
(x,Q,t_\ell).
\label{foursixteen}
\end{equation}
This relation expresses the permanence of the records of the history
$(\alpha_\ell\cdots\alpha_1)$ in the independently-evolving functions
$\chi^{k\alpha_1} \chi^{k\alpha_2\alpha_1}, \cdots$ at times later than
$t_\ell$. Translating to the Heisenberg picture, we see that
 (\ref{foursixteen})
is the general relation
(\ref{threetwentytwo}).  Thus one of the two defining conditions of
strong decoherence is satisfied in this particular model. In
the discussion of (\ref{threetwentytwo}) we mentioned that it could only
be expected to be satisfied approximately, while here it is satisfied
exactly as a consequence (\ref{foureight}). However, as we also
mentioned above the evolution, (\ref{fourthree}) can be expected to hold
only approximately, so there is in fact no conflict with the result
(\ref{foursixteen}).

To complete the demonstration that the idealized model under discussion
in this Section provides an example of strong decoherence, it remains
only to show that the first of the two defining conditions is satisfied,
namely, the condition (\ref{threenine}) or (\ref{threethirteen}) that
at time $t_k$ future extensions of history to time $t_k$
 are contained in the set of  operators $M$
effectively commuting with the records at $t_k$.  Specifically, we seek
to show
\begin{equation}
{\rm I.}\qquad\qquad\left[U^\dagger_{k\ell}
H^{(k,\ell+1)}_{\alpha_k\cdots\alpha_1}\ ,
R^\ell_{\alpha_\ell\cdots\alpha_1}\right]\,
R^\ell_{\alpha_\ell\cdots\alpha_1} |\Psi\rangle = 0\ ,
\label{fourseventeen}
\end{equation}
where the $R^\ell_{\alpha_\ell\cdots\alpha_1}$ are specified in the
Schr\"odinger picture by (\ref{foureight}), $U_{k\ell}$ is given by
(\ref{fourthirteen}), and we have introduced a convenient notation for
the extensions of histories:
\begin{mathletters}
\label{foureighteen}
\begin{eqnarray}
H^{(k, \ell+1)}_{\alpha_k\cdots\alpha_1} & \equiv &
P^k_{\alpha_k\cdots\alpha_1} P^{k-1}_{\alpha_{k-1}\cdots\alpha_1} \cdots
P^{\ell+1}_{\alpha_{\ell+1}\cdots\alpha_1}
\label{foureighteena}\\
& = & e^{iHt_k} \hat P_{\alpha_k\cdots\alpha_1} e^{-iH(t_k-t_{k-1})}
\hat P^{k-1}_{\alpha_{k-1}\cdots\alpha_1} \cdots
\hat P^{\ell+1}_{\alpha_{\ell+1}\cdots\alpha_1} e^{-iHt_{\ell+1}}
\label{foureighteenb}\\
&\equiv& e^{iHt_k} \hat H^{(k, \ell+1)}_{\alpha_k\cdots\alpha_1}
e^{-iHt_{\ell + 1}}\ .
\label{foureighteenc}
\end{eqnarray}
\end{mathletters}
With a little care, (\ref{fourseventeen}) may be transformed back to the
Schr\"odinger picture where it becomes the condition
\begin{equation}
\hat R^\ell_{\alpha_\ell\cdots\alpha_1}  e^{iH_0(t_k-t_\ell)} \hat
H^{(k, \ell)}_{\alpha_k\cdots\alpha_1} \Psi_{\alpha_\ell\cdots\alpha_1}
(x,Q,t_\ell)
=  e^{iH_0(t_k-t_\ell)} \hat H^{(k, \ell)}_{\alpha_k\cdots\alpha_1}
\Psi_{\alpha_\ell\cdots\alpha_1} (x,Q,t_\ell)\ ,
\label{fournineteen}
\end{equation}
where $\Psi_{\alpha_\ell\cdots\alpha_1} (x,Q,)$ is the branch wave
function corresponding to history $(\alpha_\ell, \cdots \alpha_1)$.
 Explicitly [{\it cf.} (\ref{henry}],
\begin{equation}
\Psi_{\alpha_\ell\cdots\alpha_1} (x,Q,t_\ell) = \hat
R^{\ell}_{\alpha_\ell\cdots\alpha_1} \Psi (x,Q,t_\ell) = \hat
H^{\ell}_{\alpha_\ell\cdots\alpha_1}\Psi(x,Q,t_0)
\label{fourtwenty}
\end{equation}
as illustrated in (\ref{fourseven}).
(These branch wave functions are not normalized as were those in (\ref{henry}).
) The operator $\hat H^{(k, \ell+1)}_{\alpha_k\cdots\alpha_{\ell+1}}$
simply extends the branch forward
to yield the following condition:
\begin{equation}
\hat R^\ell_{\alpha_\ell\cdots\alpha_1} e^{iH_0(t_k-t_\ell)}
\Psi_{\alpha_k\cdots\alpha_1} (x,Q,t_k)
 = \Psi_{\alpha_k\cdots\alpha_1} (x,Q,t_k).
\label{fourtwentyone}
\end{equation}
This can be seen to be satisfied as follows:  We have first
\begin{equation}
\Psi_{\alpha_k\cdots\alpha_1}(x,Q,t_k) = \chi^{k\alpha_1} (Q^1)
\chi^{k\alpha_2\alpha_1} (Q^2) \cdots
\chi^{k\alpha_\ell\cdots\alpha_1}(Q^\ell)
\times \left({{\rm functions\ of}\ x\ {\rm and}}
\atop {Q^{\ell+1}\cdots Q^n}\right)\ .
\label{fourtwentytwo}
\end{equation}
The crucial point is that the Hamiltonian $H_0$ was assumed to act
{\it independently} on all the co\"ordinates and the co\"ordinates $Q^1\cdots
Q^\ell$ were assumed to effectively
decouple from the rest after time $t_\ell$ .
 Thus the operator $\exp (iH_0(t_k-t_\ell))$
acts to evolve the individual $\chi$'s back to time $t_\ell$, {\it viz.}
\begin{equation}
e^{iH_0(t_k-t_\ell)}\Psi_{\alpha_k\cdots\alpha_1} (x, Q, t_k) =
\chi^{\ell\alpha_1} (Q^1) \chi^{\ell\alpha_2\alpha_1} (Q^2) \cdots
\chi^{\ell\alpha_\ell\cdots\alpha_1} (Q^1)\times \left({{\rm functions\ of}\
x\ {\rm and}} \atop {Q^{\ell+1}\cdots Q^n}\right).
\label{fourtwentythree}
\end{equation}
With this result, it is immediate that (\ref{fourtwentyone}) is
satisfied and therefore also (\ref{fourseventeen}). Thus both conditions for
strong decoherence are satisfied in this model.

A consistency condition (\ref{threeeighteen}) is  implicit in the
assumption that future histories could be drawn from the pool
$\{M\}^\ell_{\alpha_\ell\cdots\alpha_1}$ of operators that effectively
commute with the records at time $t_\ell$. That does not have to be
checked separately because we showed in the discussion
following (\ref{threefifteen}) that the condition of nesting of
records implies the necessary consistency condition, and nesting of the
records was verified explicitly for this model in eq.~(\ref{fourfourteen}).

The extensions of histories $\hat H^{(k, \ell)}_{\alpha_k\cdots\alpha_1}$ do
not exhaust the set  of
operators $\{\hat M\}^\ell_{\alpha_\ell\cdots\alpha_1}$ that effectively
commute with $\hat R^\ell_{\alpha_\ell\cdots\alpha_1}$. The
$\hat R^\ell_{\alpha^\prime_\ell\cdots\alpha^\prime_1}$ for any
$(\alpha^\prime_\ell, \cdots, \alpha^\prime_1)$ provide a simple
example of a different operator.  More generally, it is not difficult to
see that the set $\{\hat M\}^\ell_{\alpha_\ell\cdots\alpha_1}$ includes any
operator acting on ${\cal H}^x \otimes
\hat R^{\ell\perp}_{\alpha_\ell\cdots\alpha_1}\otimes {\cal H}^{Q_{\ell+1}}
\otimes \cdots \otimes {\cal H}^{Q_{\ell+1}}$, where
$\hat R^{\ell\perp}_{\alpha_\ell\cdots\alpha_1}$ is the subspace of
${\cal H}^{Q_1} \otimes \cdots \otimes {\cal H}^{Q_\ell}$ orthogonal to
the subspace defined by $\hat R^\ell_{\alpha_\ell\cdots\alpha_1}$. There may
be other operators belonging to the set
$\{\hat M\}^\ell_{\alpha_\ell\cdots\alpha_1}$.

This model provides a concrete illustration of the connection,
described in the previous Section, that
can be made between
the notion of strong decoherence and the work of many authors who
investigate the evolution of a reduced density matrix for a system in
the presence of an environment ({\it e.g.}~\cite{Zeh71,Zursum,UZ89,PZ93}).
We discussed this kind of connection in the context of similar models in
(\cite{GH93a}, Section IV) and
related ideas were discussed by Finkelstein \cite{Fin93} in connection with his
``PT-decoherence''.

Consider for definiteness the history $(\alpha_1,\alpha_2)$. Among the
operators of $\{\hat M\}^3_{\alpha_3\alpha_2\alpha_1}$ for any value of
$\alpha_3$ are those that act
Only on the Hilbert space ${\cal H}^x$. These form a  linear set
closed under Hermitian conjugation and multiplication.
The construction of a ``reduced'' matrix
(\ref{ed}) may therefore be applied, with
$\{\hat F\}^3_{\alpha_2\alpha_1}$ being the set of all operators of the
form $\hat
O^x \otimes I^Q$, where $\hat O^x$ is any operator in ${\cal
H}^x$. The $\hat A^{\alpha_2\alpha_1}_i$ are of the form
$\hat A^x_i\otimes I^Q$ where $\hat A^x_i$ is an operator basis in ${\cal H}^x$
({\it e.g.} $|x'\rangle\langle x |$).
The ``reduced'' density matrix so defined is thus nothing more than
the usual  reduced density matrix for the ``system''
described by the $x$'s.
It is constructed by tracing out the $Q$'s from the full density matrix. For
example, assuming the evolution described by eq. (\ref{fourthree}), the
reduced density matrix for the branch corresponding to the history
$(\alpha_1,\alpha_2)$ is
\begin{eqnarray}
\tilde \rho^{\alpha_2\alpha_1}(x',x)&=&
\left(\chi^{3\alpha_1} (Q^1), \chi^{3\alpha_1} (Q^1)\right)
\left(\chi^{3\alpha_2\alpha_1} (Q^2), \chi^{3\alpha_2\alpha_1}
(Q^2)\right) \\ \nonumber
&&\times\left(\chi^{3\alpha'_3\alpha_2\alpha_1}(Q^3),
\chi^{3\alpha_3\alpha_2\alpha_1}
 (Q^3)\right)  \sum\limits_{\alpha'_3}
\sum\limits_{ \alpha_3}\sum\limits_{r'_3\in\alpha'_3}
\sum\limits_{r_3\in\alpha_3}
\phi^{3\alpha_2\alpha_1}_{r'_3} (x') \phi^{3\alpha_2\alpha_1 *}_{r_3} (x)\\
\nonumber
&&
\left(\tilde\chi^{3\alpha_2\alpha_1}_{r'_3}(Q^4, Q^5,\cdots),
\tilde\chi^{3\alpha_2 \alpha_1}_{r_3} (Q^4, Q^5,\cdots)\right)\ . \label{joe}
\end{eqnarray}
(A hat should properly be on the $\tilde\rho$ but is omitted for
typographical reasons.) This reduced density matrix is easily seen to
be
diagonal in $\alpha_3$ because of the orthogonality
of the record states $\chi^{3\alpha_3\alpha_2\alpha_1}(Q^3)$ in
$\alpha_3$ [{\it cf.}(\ref{fourfour})]. Thus orthogonality of
generalized records guarantees the diagonality of the
reduced density matrix in the quantities correlated with those records.

The operators of ${\cal H}^x$ are not the only possible sets of $F$'s
in this model. They could be augmented by including, for example,
operators acting on ${\cal H}^{Q^4}\otimes{\cal H}^{Q^5}\otimes \cdots$
or on the subspace of ${\cal H}^{Q^1}$ orthogonal to
$\chi^{3\alpha_1}(Q^1)$ and similarly for the other record states.
These will result in different, less ``reduced'',  density matrices, but strong
decoherence guarantees the diagonality of them all in $\alpha_3$.

The orthogonality of the record states implies much more than just the
diagonality of the reduced density matrices that we have constructed.
For example, we have
\begin{equation}
\int dQ \Psi^3_{\alpha'_3\alpha_2\alpha_1}(x',Q)
\Psi^{*3}_{\alpha_3\alpha_2\alpha_1}(x,Q) =0, \quad
\alpha'_3 \ne  \alpha_3 \ , \label{mary}
\end{equation}
and many other relations besides. Eq.(\ref{mary}) is the general
relation (\ref{tom1}) that follows in this model because operators
on ${\cal H}^x$ are closed under multiplication.  The general content
of the diagonalization of the ``reduced'' density matrices that can
be constructed is perhaps best exhibited by relations such as this.

Thus strong decoherence implies the
diagonalization of a variety of reduced density matrices in this model but
generalizes that idea to more realistic cases where the assumptions of
the model do not apply.

The model we have been discussing expresses concretely the idea
of a physical mechanism of decoherence in which phases are dispersed
irretrievably into an ``environment'', as in the picture
elaborated with increasing precision by
the pioneers of this subject (e.g.~\cite{Hei30}, \cite{Boh34},
\cite{Zeh71}, \cite{Zursum}). However, this model is too restrictive for
realistic application. First, it is highly idealized, as we have pointed
out in several places in the discussion.  More importantly, the
realistic hydrodynamic variables that  characterize the
coarse grainings of the usual quasiclassical realm do not correspond to
a distinction between system co\"ordinates and environmental co\"ordinates. The
general notion of strong decoherence described in Section III is an
idealization that captures the physical ideas without such strong and
unrealistic assumptions as are made in this model. In the next
Section we shall discuss the role that strong decoherence can play in
defining classicality.

\section{Classicality}
\label{sec:V}

Quantum mechanics, along with the correct theory of the
elementary particles (represented by the Hamiltonian $H$) and the
correct initial condition in the universe (represented by the
state vector $|\Psi\rangle$), presumably exhibits a great many
essentially different strongly decohering realms, but only some of those
are quasiclassical.  For the quasiclassical realms to be
viewed as an emergent feature of $H$, $|\Psi\rangle$, and quantum
mechanics, a good technical definition of classicality is required. (One
can then go on to investigate whether the theory exhibits many
essentially inequivalent quasiclassical realms or whether the
usual one is nearly unique.)

In earlier papers, \cite{GH90a,GH93a,GH90b}
 we have made a number of suggestions about the
definition of classicality and it is appropriate to continue that
discussion here.  It is clear that from those earlier discussions that
classicality must be related in some way to a kind of entropy for
alternative coarse-grained histories.  We must therefore begin with an
abstract characterization of entropy and then investigate the
application to histories. An entropy $S$ is always associated with a
coarse graining, since a perfectly fine-grained version of entropy in
statistical mechanics would be conserved instead of tending to increase
with time.   Classically, if all fine-grained alternatives are
designated by $\{r\}$, with probabilities $p_r$ summing to one, that
fine-grained version of entropy would be
\begin{equation}
S_{f-g} = -\sum\limits_r p_r \log\, p_r\ ,
\label{fiveone}
\end{equation}
where $\log$  means  $\log_2$ and where, for convenience, we
have put Boltzmann's constant $k$  times $ {\rm log}_e 2$
equal to unity. A true,
coarse-grained entropy has the form
\begin{equation}
S \equiv   -\sum\limits_r \tilde p_r \log \tilde p_r\ ,
\label{fivetwo}
\end{equation}
where the probabilities $\tilde p_r$ are coarse-grained averages of the
$\{p_r\}$. A coarse graining $p_r\to\tilde p_r$ must have
certain properties (see \cite{GH96,GLpp} for more details):
\def\tp{{\tilde p}}
\begin{mathletters}
\label{fivethree}
\begin{eqnarray}
{\rm 1)} &\qquad\qquad&{\rm the}\ \{\tilde p_r\}\ {\rm are\ probabilities}
\ ,\label{fivethreea}\\
{\rm 2)} &\qquad\qquad& \skew6\tilde\tp_r = \tilde p_r\ ,
\label{fivethreeb}\\
{\rm 3)} &\qquad\qquad& -\sum\limits_r p_r\log\, \tilde p_r =
-\sum\limits_r \tilde p_r \log\, \tilde p_r\ .
\label{fivethreec}
\end{eqnarray}
\end{mathletters}
These properties are not surprising for  an averaging procedure. The
significance of the last one is easily seen if we make use of the well known
fact
that for any two sets of probabilities $\{p_r\}$ and
$\{p^\prime_r\}$ we have
\begin{equation}
-\sum\limits_r p_r \log\, p_r \leq - \sum\limits_r p_r \log
\, p^\prime_r\ .
\label{fivefour}
\end{equation}
Putting $p^\prime_r = \tilde p_r$ for each $r$ and using
(\ref{fiveone}), (\ref{fivetwo}), (\ref{fivethree}), and
(\ref{fivefour}), we obtain
\begin{equation}
S_{f-g} = - \sum\limits_r p_r \log\, p_r \leq - \sum\limits_r
p_r \log \tilde p_r = -\sum\limits_r \tilde p_r \log \tilde p_r
= S\ ,
\label{fivefive}
\end{equation}
so that $S_{f-g}$ provides a lower bound for the entropy $S$. If the
initial condition and the coarse graining are related in such away that
$S$ is initially near its lower bound, then it will tend to increase for
a period of time.  That is the way the second law of thermodynamics
comes to hold.

In order to know what nearness to the lower bound means, we should examine
the upper bound on $S$. That upper bound is achieved when all
fine-grained alternatives have equal coarse-grained probabilities
$\tilde p_r$, corresponding in statistical mechanics to something like
``equilibrium'' or infinite temperature.  Each $\tilde p_r$ is then
equal to $N^{-1}$, where $N$ (assumed finite) is the number
of fine-grained alternatives, and the maximum entropy is thus
\begin{equation}
S_{\rm max} = \log\, N\ .
\label{fivesix}
\end{equation}

The simplest example of coarse graining utilizes a grouping of the
fine-grained alternatives $\{r\}$ into exhaustive and mutually exclusive
classes $\{\alpha\}$, where a class $\alpha$ contains  $N_\alpha$
elements and has lumped probability
\begin{equation}
p_\alpha \equiv \sum\limits_{r\in\alpha} p_r\ .
\label{fiveseven}
\end{equation}
Of course we have
\begin{equation}
\sum\limits_\alpha N_\alpha = N, \quad \sum\limits_\alpha
p_\alpha =  1\ .
\label{fiveeight}
\end{equation}
The coarse-grained probabilities $\tilde p_r$ in this example are the
class averages
\begin{equation}
\tilde p_r= p_\alpha/N_\alpha\ , \quad r\in\alpha\ ,
\label{fivenine}
\end{equation}
and they clearly have the properties (\ref{fivethree}). The entropy
comes out
\begin{equation}
S = -\sum\limits_\alpha p_\alpha \log \, p_\alpha +
\sum\limits_\alpha p_\alpha \log N_\alpha\ ,
\label{fiveten}
\end{equation}
where the second term contains the familiar logarithm of the number of
fine-grained alternatives (or microstates) in a
coarse-grained alternative (or macrostate), averaged over all the
coarse-grained alternatives.

Besides entropy, it is useful to introduce the concept of
algorithmic information content (AIC) as defined some thirty years ago
by Kolmogorov, Chaitin, and Solomonoff (all working
independently).\footnote{For a discussion of
the original papers see \cite{LV93}.}
For a string of bits
$s$
and a particular universal computer $U$, the AIC of the string,
written $K_U(s)$, is the length of the shortest program that
will cause $U$ to print out the string and then halt.
The string can be used as the description of some entity $e$, down
to a given level of detail, in a given language, assuming a given amount
of knowledge and understanding of the world, encoded in a
given way into bits \cite{GM95}. The AIC of the string can then be regarded as
$K_U(e)$, the AIC of the entity so described.

We now discuss a way of approaching classicality that utilizes AIC
as well as entropy. Some authors have tried to identify AIC in a
straightforward way with complexity, and in fact AIC is often called
algorithmic complexity. However, AIC is greatest for a ``random'' string
of bits with no regularity and that hardly corresponds to what is
usually meant by complexity in ordinary parlance or in scientific
discourse. To illustrate the connections among AIC, entropy or information,
and an effective notion of complexity,
take the ensemble $\widetilde E$ consisting of a set of fine-grained
alternatives $\{r\}$ together with their coarse-grained probabilities
$\widetilde p_r$.
We can then consider both $K_U(\widetilde E)$, the AIC of the ensemble, and
$K_U(r | \widetilde E)$, which is the AIC of a particular alternative $r$ given
the ensemble. For the latter we have the well known inequality (see, for
example \cite{Ben82}):

\begin{equation}
\sum\limits_r \widetilde p_r K_U (r |\widetilde E) \geq - \sum_r
\widetilde p_r\log \widetilde p_r=S\ .
\label{fiveeleven}
\end{equation}
Moreover, it has been shown by  R. Schack \cite{Schpp}
that, for any
$U$, a slight modification $U\to U^\prime$ permits
$K_{U^\prime}(r|\widetilde E)$ to be bounded on both
sides as follows:
\begin{equation}
S+ 1 \geq \sum\limits_r \widetilde p_r K_{U^\prime} (r|\widetilde E)
\geq S,
\label{fivetwelve}
\end{equation}
so that we have
\begin{equation}
\sum\limits_r \widetilde p_r K_{U^\prime} (r|\widetilde E)\approx
S.
\label{fivethirteen}
\end{equation}
(Previous upper bounds had ${\cal O}(1)$ in place of 1, but there was
nothing to prevent ${\cal O}(1)$ from being millions or trillions of
bits!)

Looking at the entropy $S$
as a close approximation to $\sum_r
\widetilde p_r K_{U^\prime} (r | \widetilde E)$, we see that it
is natural to complete it by adding to it the quantity
$K_{U^\prime}(\widetilde E)$ --- the AIC of the {\it ensemble}
with respect to the same
universal computer $U^\prime$. This last quantity can be
connected with the idea of effective complexity --- the length of  the most
concise description of the perceived regularities of an entity $e$.
Any particular set of regularities can be expressed by describing $e$ as
a member of an ensemble $\widetilde E$ of possible entities sharing
those regularities.
Then $K_{U^\prime}(\widetilde E) $ may be identified
with the effective complexity of $e$  or of the ensemble $\widetilde E$
\cite{GLpp,GM95}. Adding this effective complexity to $S$, we have:
\begin{equation}
\Sigma \equiv K_{U^\prime }(\widetilde E) +S \ .
\label{fivethirteena}
\end{equation}
This sum of the the effective complexity and the entropy (or Shannon
information) may be labeled either ``augmented entropy'' or ``total
information''. If the coarse graining is the simple one obtained by
partitioning the set of fine-grained alternatives $\{r\}$ into classes
$\{\alpha\}$ with cardinal numbers $N_\alpha$, then the total
information becomes
\begin{equation}
\Sigma =  K_{U^\prime}(\widetilde E) -
\sum\limits_\alpha p_\alpha \log
p_\alpha + \sum\limits_\alpha p_\alpha \log N_\alpha\ .
\label{fivefourteen}
\end{equation}

In (\ref{fivethirteena}), the first term becomes smaller as the set
of perceived regularities becomes simpler, while the second term becomes
smaller as the spread of possible entities sharing those regularities is
reduced. Minimizing $\Sigma$ corresponds to optimizing the choice of
regularities and the resulting effective complexity thereby  becomes
less subjective. Thus,
the total information or augmented entropy is useful in a wide variety
of contexts \cite{GLpp,GM95}.  We apply it here to sets of  alternative
 decohering coarse-grained histories in quantum mechanics.

The general idea of augmenting entropy with a term referring to
algorithmic information  content was proposed
in a different context by Zurek \cite{Zur89}. However, as far as we
know, the emphasis on the utility of the quantity $\Sigma$ in
(\ref{fivethirteena}) and (\ref{fivefourteen}) is new.
We discussed the general idea of an entropy for histories
in \cite{GH90a}. Earlier, Lloyd and
Pagels \cite{LP88} introduced a quantity they called thermodynamic depth,
applicable to alternative coarse-grained classical histories $\alpha$.
They defined it as
\begin{equation}
D = \sum\limits_\alpha p_\alpha \log (p_\alpha/q_\alpha)\ ,
\label{fivefifteen}
\end{equation}
where $q_\alpha$ is an ``equilibrium probability'', which in our
notation would be $N_\alpha/N$ for the simple coarse graining we have
discussed.  We clearly have
\begin{equation}
D= \log N + \sum\limits_\alpha p_\alpha \log p_\alpha -
\sum\limits_\alpha p_\alpha \log N_\alpha
\label{fivesixteen}
\end{equation}
or
\begin{equation}
D=S_{\rm max} - S
\label{fiveseventeen}
\end{equation}
for the set of alternative coarse-grained histories. We see that
thermodynamic depth is intimately related to the notion of an entropy
for histories.

In applying augmented entropy to sets of coarse-grained histories in
quantum mechanics, one must take into account that there are
infinitely many different sets of fine-grained histories and that these
sets do not
generally have probabilities because they fail to decohere.
The quantities $N_\alpha$ may therefore conceivably depart from their
obvious definition as the numbers of fine-grained histories in the
coarse-grained classes $\{\alpha\}$. In fact, there may be
some latitude in the precise definition of the
complexity and entropy terms in the total information
(\ref{fivethirteena}). For example, one could consider instead of
$\widetilde E$ an ensemble
$\hat E$ consisting of the coarse-grained histories
$\alpha=(\alpha_1,\alpha_2,\cdots,\alpha_n)$, their probabilities
$p_\alpha$, and the numbers $N_\alpha$. A more general definition
of the entropy $S$ of histories  may help to define these numbers.
The generalized Jaynes construction for coarse-grained histories
 provides one framework for doing this \cite{GH90a}.
In the most general situation, such a construction defines the
entropy $S$ as the maximum of
$-Tr(\tilde \rho \log \tilde \rho)$ over all density matrices $\tilde \rho$
that
preserve the decoherence and probabilities of a given ensemble $E$ of
coarse-grained histories.  Other Jaynes-like constructions may also be useful,
for example ones that
define entropy by proceeding step by step through the histories.
We are investigating these various possibilities.

In any case,
our augmented entropy (5.15) for coarse-grained decohering histories
in quantum mechanics is a negative measure of
classicality: the smaller the quantity, the closer the set of
alternative histories is to a quasiclassical realm.
Reducing the first term in (5.15) favors making the description of the
sequences of projections simple in terms of the field variables of the
theory and the Hamiltonian $H$.
It favors sets of
projections at different times that are related to one another by time
translations, as are many sequences of projections on quasiclassical
alternatives
at different times in the usual quasiclassical realm.

Reducing the second term favors more nearly deterministic situations in
which the spread of probabilities is small. Approximate determinism
is, of course, a property of a quasiclassical realm. Reducing the last term
corresponds roughly to approaching ``maximality'', allowing the finest
graining that still permits decoherence and nearly classical behavior.
A quasiclassical realm must be maximal in order for it to be a feature
exhibited by the initial condition and Hamiltonian and not a matter of
choice by an observer.

Any proposed measure of closeness to a quasiclassical realm must be
tested by searching for pathological cases of alternative decohering
histories that make the quantity small without resembling
quasiclassical realm of everyday experience.
The worst pathology occurs for a set of histories
in which the $P$'s at every time are projections on $|\Psi\rangle$ and
on states orthogonal to $|\Psi\rangle$. We see that in this pathological
case the description of the histories and their probabilities is simple
because the description of the initial state is simple, so that
$K_{U'}(\widetilde E)$ is small. The term $-\sum_\alpha  p_\alpha {\rm log}
p_\alpha$
is zero
and the third term is also zero since the only $\alpha$ with
$p_\alpha \neq  0$ corresponds to projecting onto the pure state
$|\Psi\rangle$, so that $N_\alpha$ is one and ${\rm log}N_\alpha$
vanishes.

Evidently the smallness of $\Sigma$ is not by itself a
sufficient criterion for characterizing a quasiclassical realm.
Further criteria can be introduced if we require that
quasiclassical
realms be strongly decohering with suitable restrictions on the
sets $\{M\}_\alpha$ of operators from which the future histories are
constructed. Requiring strong decoherence ensures a physical mechanism
of decoherence and guarantees the permanence of the past. The
sets $\{M\}_\alpha$ must be restricted so as to rule out pathologies
such as discussed above.
Presumably they must all belong to a huge set with certain
straightforward properties.
Those properties might be
connected with locality, since quantum field theory is perfectly local.
(Even superstring theory is local --- although the string is an extended
object, interaction among strings is always local in spacetime.)
 It would be in this way that strong decoherence enters a definition
of classicality.

A quasiclassical realm would then be characterized in quantum
mechanics as a strongly decoherent set of histories, with suitable
restrictions on the sets $\{M\}_\alpha$, that minimizes the augmented
entropy given by (5.15). Quasiclassical realms so defined would be an
emergent feature of $H$, $|\Psi\rangle$, and quantum mechanics --- a
feature of the universe independent of human choice. In principle,
given $H$ and $|\Psi\rangle$, we could compute the quasiclassical realm
that these theories exhibit. We could then investigate the important
question of whether the usual quasiclassical realm is essentially unique
or whether the quantum mechanics of the universe exhibits essentially
 inequivalent quasiclassical
realms. Either conclusion would be of central importance for
understanding quantum mechanics.

We are continuing our efforts to complete, refine, and investigate the
consequences of this definition of classicality.

\

\acknowledgments

Much of this work was done at the Aspen Center for Physics over several
summers. The work of M.~Gell-Mann was supported in part by grants to the
Santa Fe Institute by Jeffrey Epstein, David Schiff, and Gideon
Gartner.
That of J.B.~Hartle was supported in part by the NSF grant PHY90-08502.

\end{document}